\documentclass{PoS}
\usepackage{graphicx}
\usepackage{verbatim}
\usepackage{amsmath}

\title{NLO jet production at central rapidities in $k_T$ factorization}

\ShortTitle{NLO jet production at central rapidities in $k_T$ factorization}

\author{J. Bartels\\
        II. Institut f\"ur Theoretische Physik, Universit\"at Hamburg, Luruper Chaussee 149, D-22761~Hamburg, Germany\\
        E-mail: \email{jochen.bartels@desy.de}}

\author{A. Sabio Vera\\
        Physics Department, Theory Division, CERN, CH-1211 Geneva 23, Switzerland\\
        E-mail: \email{Agustin.Sabio.Vera@cern.ch}}

\author{\speaker{F. Schwennsen}\\
        II. Institut f\"ur Theoretische Physik, Universit\"at Hamburg, Luruper Chaussee 149, D-22761~Hamburg, Germany\\
        E-mail: \email{florian.schwennsen@desy.de}}

\abstract{\noindent
In this contribution we discuss the inclusive production of jets in central 
regions of rapidity in the context of $k_T$--factorization at 
next--to--leading order (NLO). We work in the Regge limit of QCD and use 
the NLO BFKL results. A jet cone definition is proposed together with a 
phase--space separation into multi--Regge and 
quasi--multi--Regge kinematics. We discuss scattering of highly virtual 
photons, with a symmetric energy scale to separate the impact factors from 
the gluon Green's function, and hadron--hadron collisions, with a 
non--symmetric scale choice.}

\FullConference{DIFFRACTION 2006 - International Workshop on Diffraction in High-Energy Physics \\
                 September 5-10 2006\\
                 Adamantas, Milos island, Greece}

\begin{document}

\newcommand{\orders}[1]{\ensuremath{\mathcal{O}\left(s^{#1}\right)}}
\newcommand{\ordershat}[1]{\mathcal{O}\left(\hat{s}^{#1}\right)}
\newcommand{\orderslambda}[1]{\mathcal{O}\left(s_\Lambda^{#1}\right)}
\newcommand{\ordereps}[1]{\ensuremath{\mathcal{O}\left(\epsilon^{#1}\right)}}
\newcommand{\order}[1]{\mathcal{O}\left(#1\right)}
\newcommand{\dk}{d^{D-2}{\bf k}}
\newcommand{\dka}{d^{D-2}{{\bf k}_a}}\newcommand{\dkb}{d^{D-2}{{\bf k}_b}}
\newcommand{\dkl}{d^{D-2}{{\bf k}_l}}\newcommand{\dki}{d^{D-2}{{\bf k}_i}}
\newcommand{\dkone}{d^{D-2}{{\bf k}_1}}
\newcommand{\dktwo}{d^{D-2}{{\bf k}_2}}
\newcommand{\dktwoeps}{\frac{d^{D-2}{{\bf k}_2}}{\mu^{2\epsilon}(2\pi)^{D-4}}}
\newcommand{\dkjet}{d^{D-2}{{\bf k}_J}}
\newcommand{\dkjetone}{d^{D-2}{{\bf k}_{J_1}}}
\newcommand{\dkjettwo}{d^{D-2}{{\bf k}_{J_2}}}
\newcommand{\dkpure}{d^{D-2}{{\bf k}}}
\newcommand{\dkprime}{d^{D-2}{{\bf k}'}}
\newcommand{\dlambdeps}{\frac{d^{D-2}{{\bf \Lambda}}}{\mu^{2\epsilon}(2\pi)^{D-4}}}
\newcommand{\ki}{{\bf k}_i}
\newcommand{\kim}{{\bf k}_{i-1}}
\newcommand{\kip}{{\bf k}_{i+1}}
\newcommand{\kj}{{\bf k}_j}
\newcommand{\kjm}{{\bf k}_{j-1}}\newcommand{\kjp}{{\bf k}_{j+1}}
\newcommand{\km}{{\bf k}_m}
\newcommand{\kmm}{{\bf k}_{m-1}}
\newcommand{\kmp}{{\bf k}_{m+1}}
\newcommand{\kl}{{\bf k}_l}
\newcommand{\klm}{{\bf k}_{l-1}}
\newcommand{\klp}{{\bf k}_{l+1}}
\newcommand{\kn}{{\bf k}_n}
\newcommand{\knm}{{\bf k}_{n-1}}
\newcommand{\knp}{{\bf k}_{n+1}}
\newcommand{\ka}{{\bf k}_a}
\newcommand{\kb}{{\bf k}_b}
\newcommand{\kone}{{\bf k}_1}
\newcommand{\ktwo}{{\bf k}_2}
\newcommand{\kjet}{{\bf k}_J}
\newcommand{\kjetone}{{\bf k}_{J_1}}
\newcommand{\kjettwo}{{\bf k}_{J_2}}
\newcommand{\kpure}{{\bf k}}
\newcommand{\kprime}{{\bf k}'}
\newcommand{\oma}{\omega_0({\bf k}_a)}
\newcommand{\om}[1]{\omega_0(#1)}
\newcommand{\omall}{\omega_0^{LL}({\bf k}_a)}
\newcommand{\omll}[1]{\omega_0^{LL}(#1)}
\newcommand{\del}[1]{\delta^{(2)}\left(#1\right)}
\newcommand{\non}{\nonumber\\}
\newcommand{\asquare}[4]{\left|\mathcal{A}(#1,#2,#3,#4)\right|^2}
\newcommand{\bssquare}[4]{\left|\mathcal{B}_s(#1,#2,#3,#4)\right|^2}
\newcommand{\btssquare}[4]{\left|\widetilde\mathcal{B}_s(#1,#2,#3,#4)\right|^2}
\newcommand{\bsquare}[4]{\left|\mathcal{B}(#1,#2,#3,#4)\right|^2}
\newcommand{\agsquare}[4]{\left|\mathcal{A}_{2g}(#1,#2,#3,#4)\right|^2}
\newcommand{\aqsquare}[4]{\left|\mathcal{A}_{2q}(#1,#2,#3,#4)\right|^2}
\newcommand{\asbar}{\bar\alpha_s}
\newcommand{\seplog}[2]{\ln\frac{s_\Lambda}{\sqrt{#1^2 #2^2}}}
\newcommand{\qi}{{\bf q}_i}\newcommand{\qim}{{\bf q}_{i-1}}\newcommand{\qip}{{\bf q}_{i+1}}
\newcommand{\qj}{{\bf q}_j}\newcommand{\qjm}{{\bf q}_{j-1}}\newcommand{\qjp}{{\bf q}_{j+1}}
\newcommand{\ql}{{\bf q}_l}\newcommand{\qlm}{{\bf q}_{l-1}}\newcommand{\qlp}{{\bf q}_{l+1}}
\newcommand{\qn}{{\bf q}_n}\newcommand{\qnm}{{\bf q}_{n-1}}\newcommand{\qnp}{{\bf q}_{n+1}}\newcommand{\qnpp}{{\bf q}_{n+2}}
\newcommand{\qone}{{\bf q}_1}\newcommand{\qtwo}{{\bf q}_2}
\newcommand{\qa}{{\bf q}_a}\newcommand{\qb}{{\bf q}_b}
\newcommand{\qt}{\tilde{\bf q}}
\newcommand{\dqa}{d^{D-2}{{\bf q}_a}\;}\newcommand{\dqb}{d^{D-2}{{\bf q}_b}\;}
\newcommand{\dqi}{d^{D-2}{{\bf q}_i}\;}
\newcommand{\dqt}{d^{D-2}{\tilde{\bf q}}\;}
\newcommand{\omhat}{\hat{\omega}_0}\newcommand{\omhati}{\hat{\omega}_i}\newcommand{\omhatim}{\hat{\omega}_{i-1}}\newcommand{\omhatl}{\hat{\omega}_l}\newcommand{\omhatn}{\hat{\omega}_n}\newcommand{\omhatlm}{\hat{\omega}_{l-1}}

\newcommand{\shat}{\hat{s}}\newcommand{\that}{\hat{t}}\newcommand{\uhat}{\hat{u}}
\newcommand{\lambd}{{\bf{\Lambda}}}\newcommand{\delt}{{\bf{\Delta}}}

\section{Introduction}

The study of jet production in perturbative QCD is an important element of 
phenomenological studies at present and future colliders. At high energies 
the understanding of multijet events becomes mandatory. In collinear 
factorization the theoretical analysis of multijet production is complicated 
since there is a large number of contributing diagrams. However, if we 
focus on the Regge asymptotics (small--$x$ region) of scattering amplitudes
then it is possible to describe the production of a large number of jets. The 
corresponding phase space is that where the center--of--mass energy, $s$, can 
be considered asymptotically larger than any of the other scales. In this 
region the dominating Feynman diagrams are those with gluons exchanged in the 
$t$--channel. To resum contributions of the form 
$\left(\alpha_s \ln{s}\right)^n$ to all orders, with $\alpha_s$ being the 
coupling constant, it is possible to use the Balitsky--Fadin--Kuraev--Lipatov (BFKL) framework~\cite{FKL}.

The concept of a {\it Reggeized gluon} is fundamental in the construction of 
the BFKL approach. Colour octet exchange in Regge asymptotics can be described 
by a $t$--channel gluon with its propagator modified by a multiplicative 
factor depending on a power of $s$. This power corresponds to the 
{\it gluon Regge trajectory} which is a function of the transverse momenta 
and is divergent in the infrared. This divergence is removed when real 
emissions are included using gauge invariant Reggeon--Reggeon--gluon 
couplings. This allows us to describe scattering amplitudes with a large 
number of partons in the final state. The $\left(\alpha_s \ln{s}\right)^n$ 
terms correspond to the leading--order (LO) approximation and provide a simple 
picture of the underlying physics. This approximation has limitations: in leading order both 
$\alpha_s$ and the factor scaling the energy $s$ in the resummed logarithms, 
$s_0$, are free parameters not determined by the theory. These free parameters 
can be fixed if next--to--leading terms 
$\alpha_s \left(\alpha_s \ln{s}\right)^n$ are included~\cite{FLCC}. At this 
improved accuracy, diagrams contributing to the running 
of the coupling have to be included, and also $s_0$ is not longer 
undetermined. The phenomenological importance of the NLO effects has been 
recently shown for azimuthal angle decorrelations in Mueller--Navelet jets 
in Ref.~\cite{Vera:2006un}.

The LO Reggeon--Reggeon--gluon vertex corresponds to one gluon emission which 
can possibly generate a single jet. At NLO the emission vertex also contains 
Reggeon--Reggeon--gluon--gluon and Reggeon--Reggeon--quark--antiquark terms. 
In this contribution we are interested in the description of the inclusive 
production of a single jet in the NLO BFKL formalism. The relevant events will 
be those with only one jet produced in the central rapidity region of the 
detector. To find the probability of production of these events it is needed 
to introduce a jet definition in the emission vertex. This is simple at LO, 
but at NLO one should study the possibility of double emission in the 
same region of rapidity, which could lead to the production of one or two jets.

In the present text we highlight the main elements presented in the analysis 
of Ref.~\cite{Bartels:2006hg}. In that work we discuss in detail the correct 
treatment of the different scales present in the amplitudes paying particular 
attention to the separation of multi--Regge and quasi--multi--Regge 
kinematics. There we also discuss similarities and discrepancies with the 
earlier work of Ref.~\cite{Ostrovsky:1999kj}.

Our analysis is performed for two different cases: inclusive jet production in 
the scattering of two photons with large and similar virtualities, and in 
hadron--hadron collisions. In the former case the cross section has a 
factorized form in terms of photon impact factors and gluon Green's function. 
In the latter, with a momentum scale for the hadron lower than the typical 
$k_T$ entering the production vertex, the gluon Green's function needs a 
modified BFKL kernel which incorporates some $k_T$--evolution from the 
nonperturbative, and model dependent, proton impact factor to the perturbative 
jet production vertex.

For hadron--hadron scattering, our cross section formula 
contains an {\it unintegrated gluon density} which, in addition to the usual 
dependence on the longitudinal momentum fraction, typical of collinear 
factorization, carries an explicit dependence on the transverse momentum $k_T$.
 This scheme is known as $k_T$--factorization. In the small--$x$ region,
where this type of factorization has attracted particular interest, the 
BFKL framework offers the possibility to formulate, in a systematic way, the 
generalization of the $k_T$--factorization to NLO.  
It is then possible to interpret our analysis as a contribution to the 
more general question of how to formulate the unintegrated gluon density and 
the $k_T$--factorization scheme at NLO: our results can be considered as the 
small--$x$ limit of a more general formulation.

\section{Inclusive jet production at LO}

To initiate the discussion we first study the interaction between two photons 
with large virtualities $Q_{1,2}^2$ in the Regge limit 
$s \gg |t|\sim Q_1^2\sim Q_2^2$. In this region the total cross section can be 
written as a convolution of the photon impact factors with the gluon Green's 
function, {\it i.e.}
\begin{equation}
  \label{eq:total}
\sigma(s) = \int\frac{d^2 {\bf k}_a}{2\pi\ka^2}\int\frac{d^2{\bf k}_b}{2\pi\kb^2} \, 
\Phi_A(\ka) \, \Phi_B(\kb) \,\int_{\delta-i\infty}^{\delta+i\infty}\frac{d\omega}{2\pi i} \left(\frac{s}{s_0}\right)^\omega f_\omega(\ka,\kb).
\end{equation}
A convenient choice for the energy scale is $s_0=|\ka|\,|\kb|$ since this  
naturally introduces the rapidities $y_{\tilde A}$ and $y_{\tilde B}$ of the 
emitted particles with momenta $p_{\tilde A}$ and $p_{\tilde B}$ given that 
$\left(\frac{s}{s_0}\right)^\omega = e^{\omega(y_{\tilde A}-y_{\tilde B})}$.

The gluon Green's function $f_\omega$ corresponds to the solution of the BFKL 
equation
\begin{eqnarray}
  \label{eq:bfklequation}
  \omega f_\omega(\ka,\kb) &=& \delta^2(\ka-\kb)+\int d^2{\bf k}\;\mathcal{K}(\ka,\kpure)f_\omega(\kpure,\kb),
\end{eqnarray}
with kernel
\begin{eqnarray}
  \mathcal{K}(\ka,\kpure) &=& 2 \, \omega(\ka^2) \, 
\delta^{2}(\ka-\kpure) + \mathcal{K}_r(\ka,\kpure),
\end{eqnarray}
where $\omega (\ka^2)$ is the gluon Regge trajectory and $\mathcal{K}_r$ is the real 
emission contribution to the kernel which we discus in detail in the following.

\begin{center}
\begin{figure}[htbp]
  \centering
  \includegraphics[height=6cm]{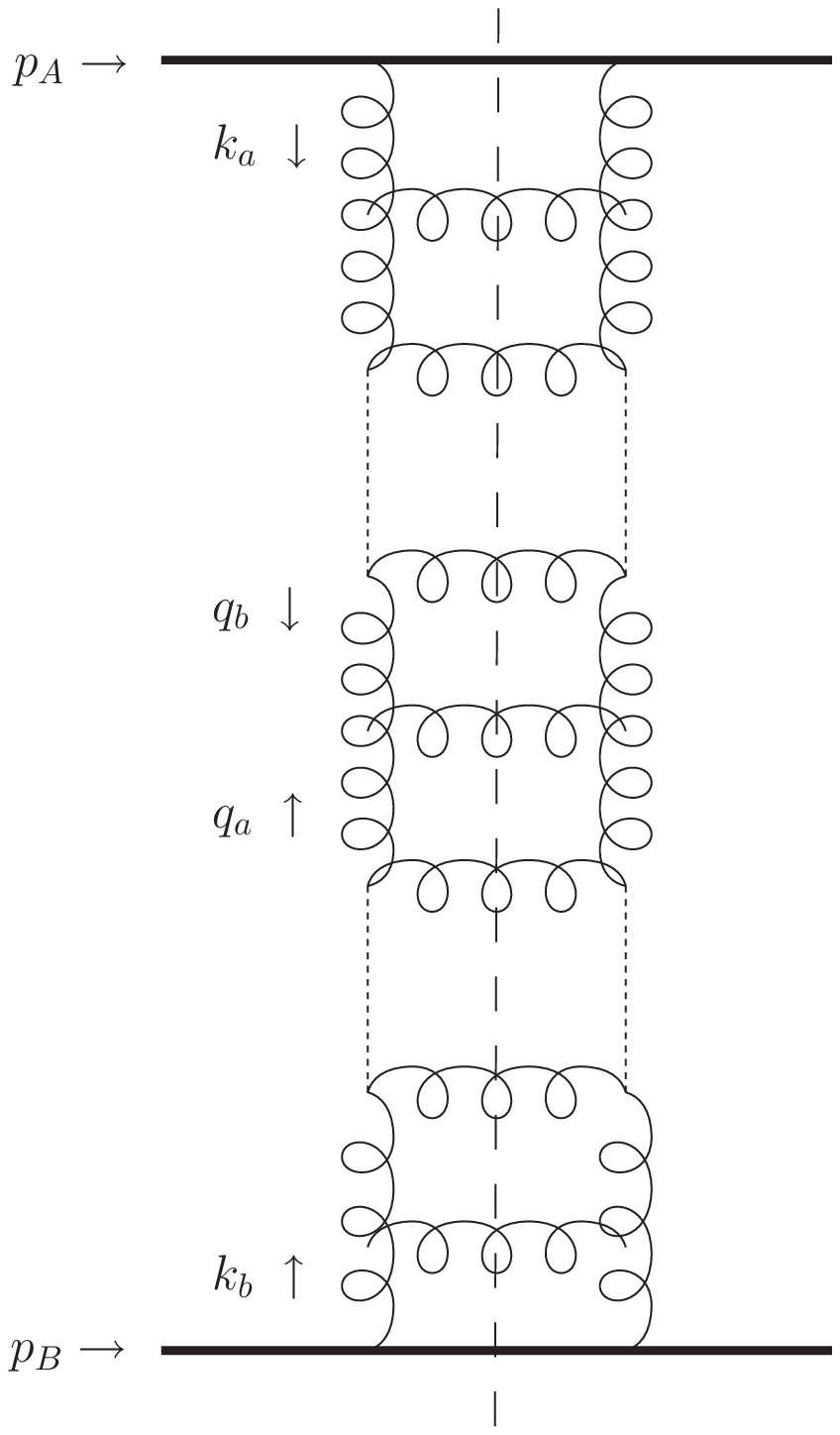}
  \hspace{2cm}
  \includegraphics[height=6cm]{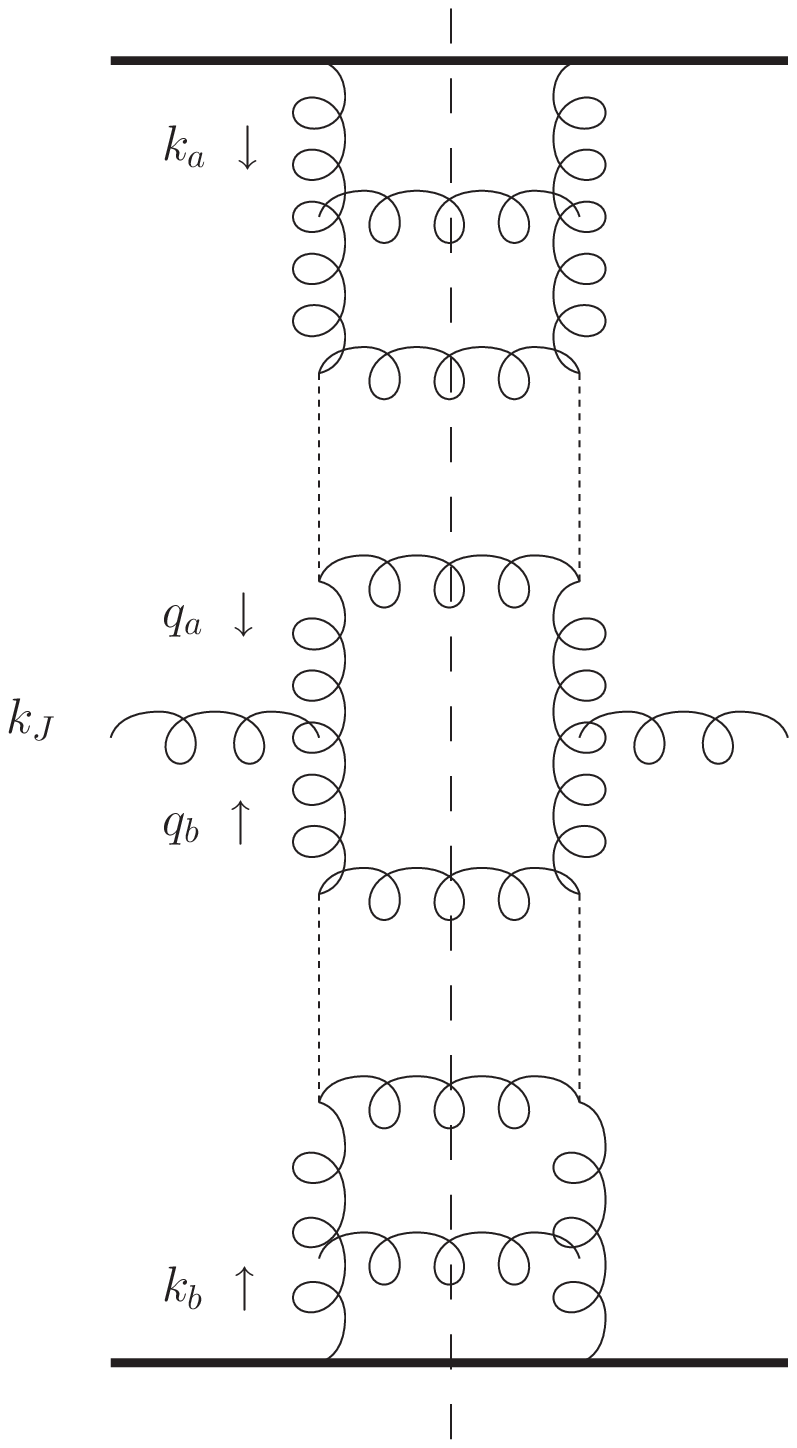}
  \caption{Total cross section and inclusive one jet production in the BFKL 
approach.}
  \label{fig:crosslo}
\end{figure}
\end{center}

It is possible to single out one gluon emission by extracting its  
emission probability from the BFKL kernel. 
By selecting one emission to be exclusive we factorize the gluon Green's 
function into two components. Each of them connects one of the external 
particles to the jet vertex, and depends on the total energies of the 
subsystems $s_{AJ} = (p_A+q_b)^2$ and $s_{BJ} = (p_B+q_a)^2$, respectively. 
We have drawn a graph indicating this separation in Fig.~\ref{fig:crosslo}. 
The symmetric situation suggests the choices $s_0^{(AJ)} =|\ka|\,|\kjet|$ and 
$s_0^{(BJ)} = |\kjet|\,|\kb|$, respectively, as the suitable energy scales for 
the subsystems. These choices can be related to the relative rapidity between 
the jet and the external particles. To set the ground for the NLO discussion 
of next section we introduce an additional integration over the rapidity 
$\eta$ of the central system in the form
\begin{multline}
\frac{d\sigma}{d^{2}\kjet dy_J} =
\int d^2 \qa \int d^2 \qb \int d\eta
\left[\int\frac{d^2 \ka}{2\pi\ka^2} \, \Phi_A(\ka) \, \int_{\delta-i\infty}^{\delta+i\infty}\frac{d\omega}{2\pi i} e^{\omega(y_A-\eta)} f_\omega(\ka,\qa)\right]\\
\times \mathcal{V}(\qa,\qb,\eta;\kjet,y_J)\;
\times \left[\int\frac{d^2 \kb}{2\pi\kb^2} \, 
\Phi_B(\kb) \int_{\delta-i\infty}^{\delta+i\infty}\frac{d\omega'}{2\pi i} e^{\omega'(\eta-y_B)}f_{\omega'}(-\qb,-\kb)\right] 
\label{eq:masterformula1}
\end{multline}
with the LO emission vertex being
\begin{equation}
\mathcal{V}(\qa,\qb,\eta;\kjet,y_J) = \mathcal{K}_r^{\rm (Born)}\left(\qa,-\qb\right) \, \del{\qa+\qb-\kjet}\,\delta(\eta-y_J). 
\label{eq:jetvertexloy}
\end{equation}

If the colliding external particles provide no perturbative scale,  as it is 
the case in hadron--hadron collisions, then the jet is the only hard scale 
in the process and we have to deal with an asymmetric situation. In such a 
configuration the 
scales $s_0$ should be chosen as $\kjet^2$ alone. At LO accuracy $s_0$ is 
arbitrary and we are indeed free to make this choice. At this stage it is 
possible to introduce the concept of {\it unintegrated gluon 
density} in the hadron. This represents the probability of resolving a gluon 
carrying a longitudinal momentum fraction $x$ from the incoming hadron, and 
with a certain transverse momentum $k_T$. Its relation to the gluon Green's 
function would be
\begin{equation}
  \label{eq:updflo}
  g(x,\kpure) = \int\frac{d^2 {\bf q}}{2\pi {\bf q}^2}\,\Phi_{P}({\bf q})\,\int_{\delta-i\infty}^{\delta+i\infty}\frac{d\omega}{2\pi i}\, x^{-\omega} f_\omega({\bf q},\kpure).
\end{equation}
With this new interpretation we can then rewrite Eq.~\eqref{eq:masterformula1} 
as
\begin{multline}
\label{eq:masterformula2}
\frac{d\sigma}{d^{2}\kjet dy_J} = \int d^2 \qa\int dx_1 \int d^2 \qb\int  dx_2\;g(x_1,\qa)g(x_2,\qb)\mathcal{V}(\qa,x_1,\qb,x_2;\kjet,y_J),
\end{multline}
with the LO jet vertex for the asymmetric situation being
\begin{multline}
\label{eq:jetvertexlo}
\mathcal{V}(\qa,x_1,\qb,x_2;\kjet,y_J)=\mathcal{K}_r^{\rm (Born)}\left(\qa,-\qb\right)\\
\times \del{\qa+\qb-\kjet}\,\delta\left(x_1-\sqrt{\frac{\kjet^2}{s}}e^{y_J}\right)\delta\left(x_2-\sqrt{\frac{\kjet^2}{s}}e^{-y_J}\right).
\end{multline}

\section{Inclusive jet production at NLO}

A similar approach remains valid when jet 
production is considered at NLO. The crucial step in this direction is 
to modify  the LO jet vertex of Eq.~\eqref{eq:jetvertexloy} and 
Eq.~\eqref{eq:jetvertexlo} to include new configurations present 
at NLO. We show how this is done in the following first subsection. 
In the second subsection we implement this vertex in a scattering process.

\subsection{The NLO jet vertex}

For those 
parts of the NLO kernel responsible for one gluon production we proceed in exactly the 
same way as at LO. The treatment of those terms related to two particle production is more complicated since  
for them it is necessary to introduce a jet algorithm. In general terms, if the two 
emissions generated by the kernel are nearby in phase space they will be considered as 
one single jet, otherwise one of them will be identified as the jet whereas the other 
will be absorbed as an untagged inclusive contribution. Hadronization effects in the 
final state are neglected and we simply define a cone of radius $R_0$ in the 
rapidity--azimuthal angle space such that two particles form a single jet if 
$R_{12} \equiv \sqrt{(\phi_1-\phi_2)^2+(y_1-y_2)^2} < R_0$. As long as only 
two emissions are involved this is equivalent to the $k_T$--clustering algorithm.

To introduce the jet definition in the $2 \rightarrow 2$ components of the 
kernel it is convenient to combine the gluon and quark matrix 
elements together with the MRK contribution:
\begin{align}
&\left({\cal K}_{Q\bar Q}+ {\cal K}_{GG}  \right) (\qa,-\qb) \equiv 
\int\dktwo\int dy_2\,\bsquare{\qa}{\qb}{\kone}{\ktwo} \nonumber\\
 =& \int\dktwo\int dy_2\,
\Bigg\{\aqsquare{\qa}{\qb}{\kone}{\ktwo}+\agsquare{\qa}{\qb}{\kone}{\ktwo}\theta(s_\Lambda-s_{12})\nonumber\\
&-{\cal K}^{\rm (Born)}(\qa,\qa-\kone) \, {\cal K}^{\rm (Born)}(\qa-\kone,-\qb)\;\frac{1}{2}\,\theta\left(\ln\frac{s_{\Lambda}}{\ktwo^2}-y_2\right)\theta\left(y_2-\ln\frac{\kone^2}{s_{\Lambda}}\right) \Bigg\}, 
\label{eq:defbsquare}
\end{align}
with ${\cal A}_{2P}$ being the two particle production amplitudes. At NLO it is necessary to separate multi-Regge kinematics (MRK) from quasi-multi-Regge kinematics (QMRK) in a distinct way. With this purpose we introduce an additional scale, $s_\Lambda$. The meaning of MRK is that the invariant mass of two emissions is considered larger than $s_\Lambda$ while in QMRK the invariant mass of one pair of these emissions is below this scale.

The NLO version of Eq.~\eqref{eq:jetvertexloy} then reads
\begin{align}
\mathcal{V}(\qa,\qb,\eta;\kjet,y_J)= & \left(\mathcal{K}_r^{\rm (Born)} +\mathcal{K}_r^{\rm (virtual)}\right)(\qa,-\qb) \Big|_{(a)}^{[y]}\non
  &\hspace{-2cm}+ \int\dktwo\;dy_2\bsquare{\qa}{\qb}{\kjet-\ktwo}{\ktwo}\theta(R_0-R_{12})\Big|_{(b)}^{[y]}\non
  &\hspace{-2cm}+ 2\int\dktwo\;dy_2\bsquare{\qa}{\qb}{\kjet}{\ktwo}\theta(R_{J2}-R_0)\Big|_{(c)}^{[y]}.\label{eq:jetvertexnloy}
\end{align}
In this expression we have introduced the notation 
\begin{align}
  &\Big|_{(a,b)}^{[y]}&&\hspace{-2.6cm}=\del{\qa+\qb-\kjet} \delta (\eta - y^{(a,b)}),  \\
  &\Big|_{(c)}^{[y]}&&\hspace{-2.6cm}=\del{\qa+\qb-\kjet-\ktwo} \delta \left(\eta-y^{(c)}\right). 
\end{align}

The various jet configurations demand several $y$ and $x$ configurations. These are related to the properties of the produced jet in different ways 
depending on the origin of the jet: if only one gluon was produced in MRK this 
corresponds to the configuration (a) in the table below, if two particles in 
QMRK form a jet then we have the case (b), and finally case (c) if the jet is 
produced out of one of the partons in QMRK. The factor of 2 in the last term 
of Eq.~\eqref{eq:jetvertexnloy} accounts for the possibility that either 
emitted particle can form the jet. The vertex can be written in a similar 
way if one chooses to work in $x$ configuration language.
Just by kinematics we get the explicit expressions for the different $x$ configurations listed in the following table:
\begin{center}
\begin{tabular}[h!]{c|c|cc}
  JET  & $y$ configurations & \multicolumn{2}{c}{$x$ configurations}
\\\hline
a) \raisebox{-1ex}{\includegraphics[height=.7cm]{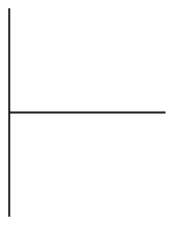}}  & 
   $y^{(a)}=y_J$ & $x_1^{(a)}=\frac{|\kjet|}{\sqrt{s}}e^{y_J}$ & 
   $x_2^{(a)}=\frac{|\kjet|}{\sqrt{s}}e^{-y_J}$ \\
b) \raisebox{-1.5ex}{\includegraphics[height=.7cm]{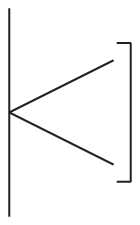}} & 
   $y^{(b)}=y_J$ & $x_1^{(b)}=\frac{\sqrt{\Sigma}}{\sqrt{s}}e^{y_J}$ & 
   $x_2^{(b)}=\frac{\sqrt{\Sigma}}{\sqrt{s}}e^{-y_J}$\\
c) \raisebox{-2ex}{\includegraphics[height=.7cm]{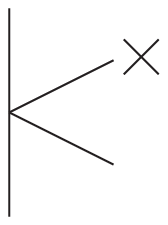}} & 
   $y^{(c)}=\frac{1}{2}\ln\frac{x_1^{(c)}}{x_2^{(c)}}$ & $ x_1^{(c)}=\frac{|\kjet|}{\sqrt{s}}e^{y_J}+\frac{|\ktwo|}{\sqrt{s}}e^{y_2}$ &   {\small $ x_2^{(c)}=\frac{|\kjet|}{\sqrt{s}}e^{-y_J}+\frac{|\ktwo|}{\sqrt{s}}e^{-y_2}$}
\end{tabular}
\end{center}


The NLO virtual correction to the one--gluon emission kernel, 
${\cal K}^{(v)}$, was originally calculated in 
Ref.~\cite{bib:kvirt}. It includes explicit infrared divergences which are canceled by the real contributions. 
The introduction of the jet definition divides the phase space into different sectors.
Only if the divergent terms belong to the same 
configuration this cancellation can be shown analytically. 
With this in mind we 
add the singular parts of the two particle production 
$|\mathcal{B}_s|^2$ in the configuration $(a)$ 
multiplied by $0=1-\theta(R_0-R_{12})-\theta(R_{12}-R_0)$:
\begin{eqnarray}
\mathcal{V} &=& \bigg[\left(\mathcal{K}_r^{\rm (Born)}+\mathcal{K}_r^{\rm (virtual)}\right)(\qa,-\qb)
+\int\dktwo\;dy_2\bssquare{\qa}{\qb}{\kjet-\ktwo}{\ktwo}\bigg] \Big|_{(a)}\nonumber\\
&+& \int\dktwo\;dy_2\bigg[\bsquare{\qa}{\qb}{\kjet-\ktwo}{\ktwo}\Big|_{(b)}
-\bssquare{\qa}{\qb}{\kjet-\ktwo}{\ktwo}\Big|_{(a)}\bigg]\theta(R_0-R_{12})\nonumber\\
&+&2\int\dktwo\;dy_2\bigg[\bsquare{\qa}{\qb}{\kjet}{\ktwo}\theta(R_{J2}-R_0)\Big|_{(c)}\nonumber\\
&&\hspace{0.4cm}-\bssquare{\qa}{\qb}{\kjet-\ktwo}{\ktwo}\theta(R_{12}-R_0)\theta(|\kone|-|\ktwo|)\Big|_{(a)}\bigg].
\label{eq:freeofsing}
\end{eqnarray}

The cancellation of divergences within the first line is now the same 
as in the calculation of the full NLO kernel.
The remainder is explicitly free 
of divergences as well since these have been subtracted out.

\subsection{Embedding of the jet vertex}

The NLO corrections to the kernel have been derived in the situation of the scattering of two objects with an intrinsic hard scale. Hence in the case of $\gamma^*\gamma^*$ scattering the equation \eqref{eq:masterformula1} is valid also at NLO if we replace the building blocks by their NLO counterparts. The most important piece being the jet vertex, which should be replaced by the one derived in the previous subsection.

We now turn to the case of hadron collisions where 
MRK has to be necessarily modified to include some evolution in the 
transverse momenta, since the momentum of the jet will be much 
larger than the typical transverse scale associated to the hadron.
In the LO case we have already explained that, in order to move from the symmetric 
case to the asymmetric one, it is needed to change the energy scale.
The independence of the result from this choice is guaranteed by a 
compensating modification of the impact factors
\begin{eqnarray}
\label{newimpactfactor}
 \widetilde{\Phi}(\ka)&=& \Phi(\ka) -\frac{1}{2}{\ka^2}\int d^2 {\bf q} 
\frac{\Phi^{\rm (Born)}({\bf q})}{{\bf q}^2}\mathcal{K}^{\rm (Born)}({\bf q},\ka)
\ln\frac{{\bf q}^2}{\ka^2}
\end{eqnarray}
and the evolution kernel
\begin{eqnarray}
\label{newkernel}
  \widetilde{\mathcal{K}}(\qone,\qtwo) &=& \mathcal{K}(\qone,\qtwo)
-\frac{1}{2}\int d^2 {\bf q} \, \mathcal{K}^{\rm (Born)}(\qone,{\bf q}) 
\, \mathcal{K}^{\rm (Born)}({\bf q},\qtwo)\ln\frac{{\bf q}^2}{\qtwo^2},
\end{eqnarray}
which corresponds to the first NLO term of a collinear 
resummation~\cite{Vera:2005jt}.

The emission vertex couples as a kind of impact factor to both  Green's functions and receives two such modifications:
\begin{eqnarray}
\label{newemissionvertex}
  \widetilde{\cal V}(\qa,\qb) &=& {\cal V}(\qa,\qb)
-\frac{1}{2}\int d^2 {\bf q} \,  
\mathcal{K}^{\rm (Born)}(\qa,{\bf q}) {\cal V}^{\rm (Born)}({\bf q},\qb)
\ln\frac{{\bf q}^2}{({\bf q}-\qb)^2}\nonumber\\
&&\hspace{1cm}-\frac{1}{2}\int d^2 {\bf q} \, {\cal V}^{\rm (Born)}(\qa,{\bf q}) \,
\mathcal{K}^{\rm (Born)}({\bf q},\qb)\ln\frac{{\bf q}^2}{(\qa-{\bf q})^2}.
\end{eqnarray}

\section{Conclusions}

In this work we have extended the NLO BFKL calculations to derive a NLO jet
production vertex in $k_T$--factorization. Our procedure was to 
`open' the BFKL kernel to introduce a jet definition at NLO in a 
consistent way.  As the central result, we have defined the jet production 
vertex and have shown how it can be used in the context of $\gamma^*\gamma^*$ 
or hadron--hadron scattering to calculate inclusive single jet cross sections. 
For this purpose we have formulated, on the basis of the NLO BFKL equation, a
NLO unintegrated gluon density valid in the small--$x$ regime.
The derived vertex can be combined with the techniques developed in 
Ref.~\cite{MC-GGF} to obtain cross sections for multijet events 
at hadron colliders.

\noindent
{\bf Acknowledgments:} A.S.V. thanks the Alexander--von--Humboldt
Foundation for financial support. F.S. is supported by the Graduiertenkolleg 
``Zuk\"unftige Entwicklungen in der Teilchenphysik''. Discussions with 
V.~Fadin and L.~Lipatov are gratefully acknowledged.

\end{document}